\documentclass[preprint,prd,aps,showpacs,showkeys,nofootinbib]{revtex4}
\usepackage{amssymb}
\usepackage{}
\usepackage{amsmath}
\usepackage{graphicx}
\usepackage{float}

\begin{document}

\preprint{}

\title{The lepton flavor violating decays of vector mesons in the MRSSM}

\author{Ke-Sheng Sun$^a$\footnote{sunkesheng@126.com;\;sunkesheng@bdu.edu.cn}, Wen-Hui Zhang$^{b,c}$\footnote{1564070519@qq.com}, Jian-Bin Chen$^{d}$\footnote{chenjianbin@tyut.edu.cn}, Hai-Bin Zhang$^{b,c}$\footnote{hbzhang@hbu.edu.cn}}

\affiliation{$^a$Department of Physics, Baoding University, Baoding, 071000,China\\
$^b$Department of Physics, Hebei University, Baoding, 071002, China\\
$^c$Key Laboratory of High-Precision Computation and Application of
Quantum Field Theory of Hebei Province, Baoding, 071002, China\\
$^d$College of Physics and Optoelectronic Engineering, Taiyuan University of Technology, Taiyuan 030024, China}
\begin{abstract}
Recently several new bounds of the lepton flavor violating decays of the neutral vector mesons $J/\psi$ and $\Upsilon(nS)$ are reported from the experiments. In the work, we analyze these processes in the scenario of the minimal R-symmetric supersymmetric standard model by means of the effective Lagrangian method. The predicted branching ratios are affected by the mass insertion parameters $\delta^{ij}$ and the contributions from different parts are comparable. Taking account of the constraints on the mass insertion parameters from the experiments, the branching ratios for the most promising processes $\Upsilon(nS)\rightarrow l\tau$ are predicted to be ten orders of magnitude smaller than the present experimental bounds.

\end{abstract}



\maketitle

\section{Introduction}
\indent\indent

Recently the Belle collaboration reported no significant signal in their search for the lepton flavor violating (LFV) decays of the neutral vector meson $\Upsilon(1S)$ \cite{Belle2022}. Using the 158 million $\Upsilon(2S)$ sample collected by the Belle detector at the KEKB collider, the upper limit of the branching ratio (BR) of $\Upsilon(1S)\rightarrow \mu\tau$ is estimated to be $2.7\times 10^{-6}$, which is 2.3 times more stringent than the previous result \cite{CLEO2008}. The collaboration also performed the first search for $\Upsilon(1S)\rightarrow e\mu$ and $\Upsilon(1S)\rightarrow e\tau$ and the estimated upper limits are $3.9\times 10^{-7}$ and $2.7\times 10^{-6}$, respectively. The BaBar collaboration reported the first search for the LFV decay $\Upsilon(3S)\rightarrow e\mu$ in a sample of 118 million $\Upsilon(3S)$ mesons from 27 fb$^{-1}$ of data collected with the BABAR detector at the SLAC PEP-II $e^+e^-$ collider \cite{BABAR2022}. The upper limit of BR$(\Upsilon(3S)\rightarrow e\mu)$ is set by $3.6\times 10^{-7}$ at 90\% confidence level. The BESIII collaboration performed the search for the LFV decays of the vector meson $J/\psi$ based on 10 billion $J/\psi$ events collected with the BESIII detector \cite{BES2103}. The upper limit of BR$(J/\psi\rightarrow e\tau)$ is determined to be $7.5\times 10^{-8}$ at the 90\% confidence level. This improves the previously published limit, which is given to be $8.3\times 10^{-6}$ in 2004 \cite{BES04598}, by two orders of magnitude. A summary of current bounds of the LFV decays of the neutral vector meson $V$ ($V$ = $J/\psi$, $\Upsilon(nS)$, $\phi$) is given in Table \ref{Tab1}. 
\begin{table}[h]
\caption{Current experimental bounds of the LFV decays of the neutral vector mesons. }
\begin{tabular}{@{}cccccc@{}} \colrule
Decay&Bound&Experiment&Decay&Bound&Experiment\\
\colrule
$J/\psi\rightarrow e\mu$&$1.6\times 10^{-7}$&BESIII (2013)\cite{BES2013}&$J/\psi\rightarrow e\tau$&$7.5\times 10^{-8}$&BESIII (2021)\cite{BES2103}\\
$J/\psi\rightarrow \mu\tau$&$2.0\times 10^{-6}$&BES (2004)\cite{BES04598}&$\Upsilon(1S)\rightarrow e\mu$&$3.9\times 10^{-7}$&BELLE (2022)\cite{Belle2022}\\
$\Upsilon(1S)\rightarrow e\tau$&$2.7\times 10^{-6}$&BELLE (2022)\cite{Belle2022}&$\Upsilon(1S)\rightarrow \mu\tau$&$2.7\times 10^{-6}$&BELLE (2022)\cite{Belle2022}\\
$\Upsilon(2S)\rightarrow e\tau$&$3.2\times 10^{-6}$&BABAR (2010)\cite{BABAR2010}&$\Upsilon(2S)\rightarrow \mu\tau$&$3.3\times 10^{-6}$&BABAR (2010)\cite{BABAR2010}\\
$\Upsilon(3S)\rightarrow e\mu$&$3.6\times 10^{-7}$&BABAR (2022)\cite{BABAR2022}&$\Upsilon(3S)\rightarrow e\tau$&$4.2\times 10^{-6}$&BABAR (2010)\cite{BABAR2010}\\
$\Upsilon(3S)\rightarrow \mu\tau$&$3.1\times 10^{-6}$&BABAR (2010)\cite{BABAR2010}&$\phi\rightarrow e\mu$&$2\times 10^{-6}$&SND (2010)\cite{SND2010}\\
\colrule
\end{tabular}
\label{Tab1}
\end{table}

It is known that, in the standard model (SM) into which neutrino oscillations (masses and mixing) are incorporated, the LFV decays (e.g., $l_1\rightarrow l_2\gamma$, $l_1\rightarrow 3l_2$, $\mu-e$ conversion, $h\rightarrow l_1 \bar{l}_2$, $\tau \rightarrow Pl_1$, where $l_{1,2}=e,\mu,\tau$) are highly suppressed by small masses of neutrinos. Under the assumption that only three massive neutrinos are present, the theoretical branching ratios of the LFV decays of the vector meson $V$ in the SM are smaller than $ 10^{-50}$ \cite{Abada113013}, which  fall out the reach of the current experiment. Nevertheless, the branching ratios of the LFV decays can be enhanced in various extensions of the SM, such as the grand unified models \cite{Pati1974,Georgi1974,Langacker1981}, the supersymmetric models with and without R-parity \cite{Haber1985,Chang2000}, the left-right symmetry models \cite{Mohapatra19751,Mohapatra19752,Senjanovic1975} and so on. The LFV decays of the neutral vector mesons have been studied in several new physics scenarios \cite{Abada113013,Kss1250172,Kss486,Huo2003,Wei2009,Dong2018,Yue2016} and in a model-independent way \cite{Nussinov2001, Gutsche2011, Gutsche2010, Hazard2016, Angelescu, Gonzalez2101}. It has been shown that BR$(V\rightarrow l_1 \bar{l}_2)$ can be significantly enhanced close to the current or future experimental sensitivities.

In this work, we will study the LFV decays $V\rightarrow l_1 \bar{l}_2$ in the minimal R-symmetric supersymmetric standard model (MRSSM) \cite{Kribs}. The MRSSM has the same gauge symmetry $SU(3)_C\times SU(2)_L\times U(1)_Y$ as the SM and the MSSM. It contains a continuous R-symmetry \cite{Fayet,Salam}, Dirac gauginos and an N=2 sector. R-symmetry forbids Majorana gaugino masses, $\mu$ term, $A$ terms and all the left-right squark and slepton mass mixings. There is a well-known tan$\beta$-enhancement in the MSSM for the anomalous magnetic dipole moment $a_\mu$ \cite{Stockinger2007,Moroi1996} and a similar enhancement for the LFV processes $\mu\rightarrow e \gamma$ and $\mu-e$ conversion \cite{Hisano1996}. However, the tan$\beta$-enhancement for these observables does not exist in the MRSSM since the absence of Majorana gaugino masses and $\mu$ term \cite{Kotlarski}. The LFV decays only originate from the off-diagonal entries in the slepton mass matrices $m_l^2$ and $m_r^2$. More studies on the phenomenology in the MRSSM can be found in literatures \cite{Diessner2014, Diessner2015, Diessner2016, Diessner2017, Diessner2019, Diessner20192, Kotlarski, Kumar, Blechman, Kribs1, Frugiuele, Jan, Chakraborty, Braathen, Athron, Alvarado, kss2020, kss20202}.

In Section \ref{sec2}, we provide a brief introduction to the MRSSM and present the notation and conventions for the operators and their corresponding Wilson coefficients. The numerical analysis is presented in Section \ref{sec3}, and Section \ref{sec4} is devoted to a conclusion.

\section{The MRSSM\label{sec2}}

In this section, we first provide a simple overview of the MRSSM. The general form of the superpotential of the MRSSM is given by \cite{Diessner2014},
\begin{eqnarray}
\mathcal{W}_{MRSSM} &=& \mu_d\hat{R}_d\cdot \hat{H}_d+\mu_u \hat{R}_u\cdot \hat{H}_u+\Lambda_d \hat{R}_d\cdot \hat{T}\hat{H}_d+\Lambda_u \hat{R}_u\cdot\hat{T}\hat{H}_u\nonumber\\
&+&\lambda_d\hat{S}\hat{R}_d\cdot\hat{H}_d+\lambda_u\hat{S} \hat{R}_u\cdot\hat{H}_u-Y_d\hat{d}\hat{q}\cdot\hat{H}_d-Y_e\hat{e}\hat{l}\cdot\hat{H}_d+Y_u\hat{u}\hat{q}\cdot\hat{H}_u,
\end{eqnarray}
where $\hat{H}_u$ and $\hat{H}_d$ are the MSSM-like Higgs weak iso-doublets, $\hat{R}_u$ and $\hat{R}_d$ are the $R$-charged Higgs $SU(2)_L$ doublets and the corresponding Dirac higgsino mass parameters are denoted as $\mu_u$ and $\mu_d$. $Y_e$, $Y_u$ and $Y_d$ are the Yukawa couplings of charged lepton, up type quark and down type quark, respectively. $\lambda_u$, $\lambda_d$, $\Lambda_u$ and $\Lambda_d$ are parameters of Yukawa-like trilinear terms involving the singlet $\hat{S}$ and the triplet $\hat{T}$, which is given by
\begin{equation}
\hat{T} = \left(
\begin{array}{cc}
\hat{T}^0/\sqrt{2} &\hat{T}^+ \\
\hat{T}^-  &-\hat{T}^0/\sqrt{2}\end{array}
\right).\nonumber
 \end{equation}

For the phenomenological studies we take the soft breaking scalar mass terms
\begin{eqnarray}
V_{SB,S} &=& m^2_{H_d}(|H^0_d|^2+|H^{-}_d|^2)+m^2_{H_u}(|H^0_u|^2+|H^{+}_u|^2)+(B_{\mu}(H^-_dH^+_u-H^0_dH^0_u)+h.c.)\nonumber\\
&+&m^2_{R_d}(|R^0_d|^2+|R^{+}_d|^2)+m^2_{R_u}(|R^0_u|^2+|R^{-}_u|^2)+m^2_T(|T^0|^2+|T^-|^2+|T^+|^2)\nonumber\\
&+&m^2_S|S|^2+ m^2_O|O^2|+\tilde{d}^*_{L,i} m_{q,{i j}}^{2} \tilde{d}_{L,j} +\tilde{d}^*_{R,i} m_{d,{i j}}^{2} \tilde{d}_{R,j}+\tilde{u}^*_{L,i}  m_{q,{i j}}^{2} \tilde{u}_{L,j}\nonumber\\
&+&\tilde{u}^*_{R,i}  m_{u,{i j}}^{2} \tilde{u}_{R,j}+\tilde{e}^*_{L,i} m_{l,{i j}}^{2} \tilde{e}_{L,j}+\tilde{e}^*_{R,{i}} m_{r,{i j}}^{2} \tilde{e}_{R,{j}} +\tilde{\nu}^*_{L,i} m_{l,{i j}}^{2} \tilde{\nu}_{L,j}.
\end{eqnarray}
All trilinear scalar couplings involving Higgs bosons to squarks and sleptons are forbidden due to the R-symmetry. The soft-breaking terms, which describe the Dirac mass terms for the gauginos and the interaction terms between the adjoint scalars and the auxiliary D-fields of the corresponding gauge multiplet, take the form
\begin{equation}
V_{SB,DG}=M^B_D(\tilde{B}\tilde{S}-\sqrt{2}\mathcal{D}_B S)
+M^W_D(\tilde{W}^a\tilde{T}^a-\sqrt{2}\mathcal{D}_W^a T^a)
+M^O_D(\tilde{g}\tilde{O}-\sqrt{2}\mathcal{D}_g^a O^a)+h.c.,
\label{}
\end{equation}
where $\tilde{B}$, $\tilde{W}$ and $\tilde{g}$ are usually MSSM Weyl fermions, $M^B_D$, $M^W_D$ and $M^O_D$ are the mass of bino, wino and gluino, respctively.

The number of neutralino degrees of freedom in the MRSSM is doubled compared to the MSSM as the neutralinos are Dirac-type. The neutralino mass matrix and the diagonalization procedure are
\begin{eqnarray}
m_{\chi^0} &=& \left(
\begin{array}{cccc}
M^{B}_D &0 &-\frac{1}{2} g_1 v_d  &\frac{1}{2} g_1 v_u \\
0 &M^{W}_D &\frac{1}{2} g_2 v_d  &-\frac{1}{2} g_2 v_u \\
- \frac{1}{\sqrt{2}} \lambda_d v_d  &-\frac{1}{2} \Lambda_d v_d  &-\mu_d^{eff,+}&0\\
\frac{1}{\sqrt{2}} \lambda_u v_u  &-\frac{1}{2} \Lambda_u v_u  &0 &\mu_u^{eff,-}\end{array}
\right),(N^{1})^{\ast} m_{\chi^0} (N^{2})^{\dagger} =m_{\chi^0}^{\textup{diag}}.
\end{eqnarray}
where the modified $\mu_i$ parameters $\mu_{\{u,d\}}^{eff,\pm}= \pm\frac{1}{2} \Lambda_{\{u,d\}} v_T  + \frac{1}{\sqrt{2}} \lambda_{\{u,d\}} v_S  + \mu_{\{u,d\}}$. $g_1$ and $g_2$ are coupling constants for the $U(1)_Y$ part and the $SU(2)_L$ part. $v_u$ and $v_d$ are the nonzero vacuum expectation values of two Higgs doublets and $\tan\beta$=$\frac{v_u}{v_d}$ is defined.

The number of chargino degrees of freedom in the MRSSM is also doubled compared to the MSSM and these charginos can be grouped to two separated chargino sectors according to their R-charge. The $\chi$-chargino sector has R-charge 1 electric charge; the $\rho$-chargino sector has R-charge -1 electric charge. Here, we do not discuss the $\rho$-chargino sector in detail since it does not contribute to the LFV decays. The $\chi$-chargino mass matrix and the diagonalization procedure are
\begin{equation}
m_{\chi^{+}} = \left(
\begin{array}{cc}
g_2 v_T  + M^{W}_D &\frac{1}{\sqrt{2}} \Lambda_d v_d \\
\frac{1}{\sqrt{2}} g_2 v_d  &\mu_d^{eff,-}\end{array}
\right),(U^{1})^{\ast} m_{\chi^{+}} (V^{1})^{\dagger} =m_{\chi^{+}}^{\textup{diag}}.
\end{equation}

The slepton mass matrix and the diagonalization procedure are
\begin{equation}
m^2_{\tilde{e}} = \left(
\begin{array}{cc}
(m^2_{\tilde{e}})_{LL} &0 \\
0  &(m^2_{\tilde{e}})_{RR}\end{array}
\right),Z^E m^2_{\tilde{e}} (Z^{E})^{\dagger} =m^{2,\textup{diag}}_{\tilde{e}}
\end{equation}
where
\begin{align}
(m^2_{\tilde{e}})_{LL} &=m_l^2+ \frac{1}{2} v_{d}^{2} |Y_{e}|^2 +\frac{1}{8}(g_1^2-g_2^2)(v_{d}^{2}- v_{u}^{2}) -g_1 v_S M_D^B-g_2v_TM_D^W \nonumber\\
(m^2_{\tilde{e}})_{RR} &= m_r^2+\frac{v_d^2}{2}|Y_e|^2+\frac{1}{4}g_1^2( v_{u}^{2}- v_{d}^{2})+2g_1v_SM_D^B.\nonumber
\end{align}
One can see that the left-right slepton mass mixing is absent in the MRSSM, whereas the A terms are present in the MSSM. The sneutrino mass matrix and the diagonalization procedure are
\begin{equation}
m^2_{\tilde{\nu}} =
\begin{array}{c}m_l^2+\frac{1}{8}(g_1^2+g_2^2)( v_{d}^{2}- v_{u}^{2})+g_2 v_T M^{W}_D-g_1 v_S M^{B}_D,Z^V m^2_{\tilde{\nu}} (Z^{V})^{\dagger} = m^{2,\textup{diag}}_{\tilde{\nu}}
\end{array}
\end{equation}
where the last two terms are newly introduced in the MRSSM.

The mass matrix for up squarks and down squarks, and the relevant diagonalization procedure are
\begin{equation}
\begin{array}{l}
m^2_{\tilde{u}} = \left(
\begin{array}{cc}
(m^2_{\tilde{u}})_{LL} &0 \\
0  &(m^2_{\tilde{u}})_{RR}\end{array}
\right), Z^U m^2_{\tilde{u}} (Z^{U})^{\dagger} =m^{2,\textup{diag}}_{\tilde{u}}, \\
m^2_{\tilde{d}} = \left(
\begin{array}{cc}
(m^2_{\tilde{d}})_{LL} &0 \\
0  &(m^2_{\tilde{d}})_{RR}\end{array}
\right),Z^D m^2_{\tilde{d}} (Z^{D})^{\dagger} =m^{2,\textup{diag}}_{\tilde{d}},\label{sud}
\end{array}
\end{equation}
where
\begin{align}
(m^2_{\tilde{u}})_{LL} &=m_{\tilde{q}}^2+ \frac{1}{2} v_{u}^{2} |Y_{u}|^2
+\frac{1}{24}(g_1^2-3g_2^2)(v_{u}^{2}- v_{d}^{2}) +\frac{1}{3}g_1 v_S M_D^B+g_2v_TM_D^W ,\nonumber\\
(m^2_{\tilde{u}})_{RR} &= m_{\tilde{u}}^2+\frac{1}{2}v_u^2|Y_u|^2+\frac{1}{6}g_1^2( v_{d}^{2}- v_{u}^{2})-\frac{4}{3} g_1v_SM_D^B,\nonumber\\
(m^2_{\tilde{d}})_{LL} &=m_{\tilde{q}}^2+ \frac{1}{2} v_{d}^{2} |Y_{d}|^2
+\frac{1}{24}(g_1^2+3g_2^2)(v_{u}^{2}- v_{d}^{2}) +\frac{1}{3}g_1 v_S M_D^B-g_2v_TM_D^W ,\nonumber\\
(m^2_{\tilde{d}})_{RR} &= m_{\tilde{d}}^2+\frac{1}{2}v_d^2|Y_d|^2+\frac{1}{12}g_1^2( v_{u}^{2}- v_{d}^{2})+\frac{2}{3} g_1v_SM_D^B.\nonumber
\end{align}

%%%%%%%%%%%%%%%%%%%%%%%%%%%%%%%%%%%%%%%%%%%%%%%%%%%%%%%%%%%%%%%%%%%
\begin{figure}[htbp]
\setlength{\unitlength}{1mm}
\centering
\begin{minipage}[c]{1\textwidth}
\includegraphics[width=6.0in]{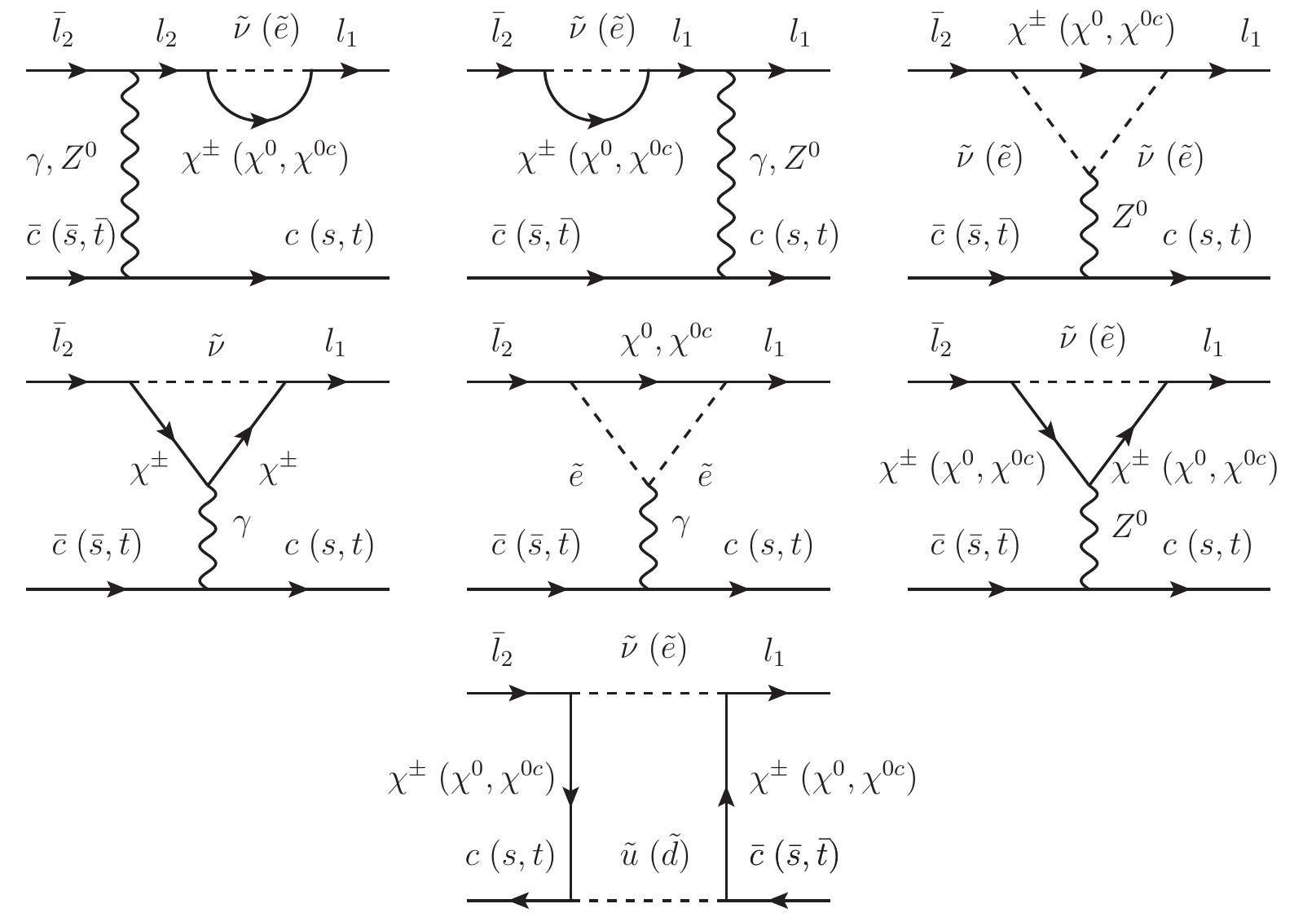}
\end{minipage}
\caption[]{Feynman diagrams contributing to $V\rightarrow l_1 \bar{l}_2$ in the MRSSM.}
\label{diag}
\end{figure}
%%%%%%%%%%%%%%%%%%%%%%%%%%%%%%%%%%%%%%%%%%%%%%%%%%%%%%%%%%%%%%%%%%%

We now focus on the LFV processes $V\rightarrow l_1 \bar{l}_2$. The relevant Feynman diagrams contributing to the LFV decays $V\rightarrow l_1 \bar{l}_2$ in the MRSSM are presented in Fig.\ref{diag}. Using the effective Lagrangian method, we present the analytical expression for the branching ratio of $V\rightarrow l_1 \bar{l}_2$. In the effective Lagrangian method, one can derive the effective Lagrangian relevant for $V\rightarrow l_1 \bar{l}_2$ as \cite{Hazard2016}
\begin{eqnarray}
-\mathcal{L}_{eff} &=&m_2(C^{l_1l_2}_{DR}\bar{l}_1\sigma^{\mu\nu}P_Ll_2
+C^{l_1l_2}_{DL}\bar{l}_1\sigma^{\mu\nu}P_Rl_2)F_{\mu\nu}+h.c.\nonumber\\
&&+\sum_q\Big[\big(C^{ql_1l_2}_{VR}\bar{l}_1\gamma^{\mu}P_R l_2+C^{ql_1l_2}_{VL}\bar{l}_1\gamma^{\mu}P_L l_2\big)\bar{q}\gamma_{\mu}q\nonumber\\
&&+\big(C^{ql_1l_2}_{AR}\bar{l}_1\gamma^{\mu}P_R l_2+C^{ql_1l_2}_{AL}\bar{l}_1\gamma^{\mu}P_L l_2\big)\bar{q}\gamma_{\mu}\gamma_5 q\nonumber\\
&&+m_2 m_q G_F \big(C^{ql_1l_2}_{SR}\bar{l}_1P_L l_2+C^{ql_1l_2}_{SL}\bar{l}_1P_R l_2\big)\bar{q}q\nonumber\\
&&+m_2 m_q G_F \big(C^{ql_1l_2}_{PR}\bar{l}_1P_L l_2+C^{ql_1l_2}_{PL}\bar{l}_1P_R l_2\big)\bar{q}\gamma_5 q\nonumber\\
&&+m_2 m_q G_F \big(C^{ql_1l_2}_{TR}\bar{l}_1\sigma^{\mu\nu}P_L l_2+C^{ql_1l_2}_{TL}\bar{l}_1\sigma^{\mu\nu}P_R l_2\big)\bar{q}\sigma_{\mu\nu}q+h.c.\Big],
\label{Leff}
\end{eqnarray}
where $P_{L/R}$=$\frac{1}{2}(1\mp\gamma_5)$ is the left (right) chiral projection operator, $G_F$ is the Fermi constant and $m_q$ is the mass of the quark $q$. The first row in Eq.(\ref{Leff}) is the dipole part which are tightly constrained by the radiative LFV decays (e.g., $l_1\rightarrow l_2\gamma$). The remains are the four-fermion dimension six lepton-quark Lagrangian where the coefficients do not include  photonic contributions but Z boson and scalar ones. 

The most general expression for the $V\rightarrow l_1 \bar{l}_2$ decay amplitude can be written as
\begin{eqnarray}
\mathcal{A}(V\rightarrow l_1 \bar{l}_2) &=&\bar{u}(p_1,s_1)\Big[A^{l_1l_2}_{V}\gamma^{\mu}+B^{l_1l_2}_{V}\gamma^{\mu}\gamma^{5}
+\frac{C^{l_1l_2}_{V}}{m_V}(p_2-p_1)_{\mu}\nonumber\\
&&-\frac{ D^{l_1l_2}_{V}}{m_V}(p_2-p_1)_{\mu}\gamma^{5}\Big]\upsilon(p_2,s_2)\epsilon^\mu(p),
\label{Amp}
\end{eqnarray}
where $m_V$ is the mass of the vector meson $V$, $\epsilon^\mu(p)$ is the polarization vector and $p$ is its momentum. The coefficients $A^{l_1l_2}_{V}$, $B^{l_1l_2}_{V}$, $C^{l_1l_2}_{V}$ and $D^{l_1l_2}_{V}$ are dimensionless constants which depend on the Wilson coefficients in Eq.(\ref{Leff}) as well as on hadronic effects associated with meson-to-vacuum matrix elements or decay constants. The coefficients in Eq.(\ref{Amp}) are given as
\begin{eqnarray}
A^{l_1l_2}_{V}&=&\sqrt{4\pi\alpha}Q_qy^2(C^{l_1l_2}_{DL}+C^{l_1l_2}_{DR})
+\kappa_V(C^{ql_1l_2}_{VL}+C^{ql_1l_2}_{VR})\nonumber\\
&&+2y^2\kappa_V\frac{f^T_V}{f_V}G_Fm_Vm_q(C^{ql_1l_2}_{TL}+C^{ql_1l_2}_{TR}),
\nonumber\\
B^{l_1l_2}_{V}&=&-\sqrt{4\pi\alpha}Q_qy^2(C^{l_1l_2}_{DL}-C^{l_1l_2}_{DR})
-\kappa_V(C^{ql_1l_2}_{VL}-C^{ql_1l_2}_{VR})\nonumber\\
&&-2y^2\kappa_V\frac{f^T_V}{f_V}G_Fm_Vm_q(C^{ql_1l_2}_{TL}-C^{ql_1l_2}_{TR}),
\nonumber\\
C^{l_1l_2}_{V}&=&\sqrt{4\pi\alpha}Q_qy(C^{l_1l_2}_{DL}+C^{l_1l_2}_{DR})
+2\kappa_V\frac{f^T_V}{f_V}G_Fm_2m_q(C^{ql_1l_2}_{TL}+C^{ql_1l_2}_{TR}),
\nonumber\\
D^{l_1l_2}_{V}&=&-\sqrt{4\pi\alpha}Q_qy(C^{l_1l_2}_{DL}-C^{l_1l_2}_{DR})
+2\kappa_V\frac{f^T_V}{f_V}G_Fm_2m_q(C^{ql_1l_2}_{TL}-C^{ql_1l_2}_{TR}),
\label{ABCD}
\end{eqnarray}
where $\alpha$ is the fine structure constant, $Q_q$ is the charge of the quark $q$, $y$ =$\frac{m_2}{m_V}$ and $\kappa_V$=1/2 is a constant for pure $q\bar{q}$ states.    From Eq.(\ref{ABCD}) we see that the contribution from the Higgs mediated self-energies and penguin diagrams is nonexistent since these diagrams only contribute to the Wilson coefficients corresponding to the scalar and the pseudo-scalar operators in Eq.(\ref{Leff}). The contribution from the tensor operators can be neglected in the MRSSM since the coefficients $C^{ql_1l_2}_{TL/TR}$ calculated from the Feynman diagrams in Fig.\ref{diag} are zero. The decay constant $f_V$ and the transverse decay constant $f^T_V$ are defined as
\begin{eqnarray}
\left\langle0\right|\bar{q}\gamma^\mu q \left|V(p)\right\rangle &=&f_Vm_V\epsilon^\mu(p), \nonumber\\
\left\langle0\right|\bar{q}\sigma^{\mu\nu} q\left|V(p)\right\rangle &=&i f^T_V (\epsilon^\mu p^\nu-\epsilon^\nu p^\mu). 
\label{decayconstant}
\end{eqnarray}

Then the amplitude in Eq.(\ref{Amp}) leads to the branching ratio BR($V\rightarrow l_1 \bar{l}_2$), which is convenient to represent
\begin{eqnarray}
{\rm BR}(V\rightarrow l_1 \bar{l}_2) &=&{\rm BR}(V\rightarrow e e)\times \big(\frac{m^2_V(1-y^2)^2}{4\pi\alpha Q_q}\big)^2\times\Big[|A^{l_1l_2}_{V}|^2+|B^{l_1l_2}_{V}|^2\nonumber\\
&&+(\frac{1}{2}-y^2)\big(|C^{l_1l_2}_{V}|^2+|D^{l_1l_2}_{V}|^2\big)
+y {\rm Re}(A^{l_1l_2}_{V}C^{l_1l_2\ast}_{V}- B^{l_1l_2}_{V}D^{l_1l_2\ast}_{V})\Big],
\label{BR}
\end{eqnarray}
where the mass of the lighter of the two leptons is neglected.

\section{Numerical analysis\label{sec3}}
\indent\indent

We carried out the calculation by using the spectrum generator SPheno-4.0.5  \cite{SPheno1,SPheno2}, where the model implementations are generated by the public code SARAH-4.14.5 \cite{SARAH, SARAH1, SARAH2,Flavor,Flavor2}. The generic expressions for the Wilson coefficients in Eq.(\ref{ABCD}) is derived with the help of the package PreSARAH-1.0.3. The explicit expressions for the Wilson coefficients in the MRSSM is obtained by adapting the generic expressions to the specific details of the MRSSM by SARAH. The Fortran code of the branching ratio in Eq.(\ref{BR}) is written by authors and this code is used by SARAH to generate the Fortran modules for SPheno. The numerical calculation of the branching ratio is done by SPheno. Note that the conventions of Ref.\cite{Hazard2016} are different from those presented in Ref.\cite{Flavor}. The Wilson coefficients are related by
\begin{eqnarray}
C^{l_1l_2}_{DR}&=&\frac{e}{2} K^L_2,\;C^{l_1l_2}_{DL}=\frac{e}{2} K^R_2,\;C^{ql_1l_2}_{VR}=B^V_{RL}+B^V_{RR},\;C^{ql_1l_2}_{VL}=B^V_{LL}+B^V_{LR}\nonumber\\
&&m_2m_qG_FC^{ql_1l_2}_{TR}=B^T_{LL}+B^T_{LR},\;m_2m_qG_FC^{ql_1l_2}_{TL}=B^T_{RL}+B^T_{RR}.
\label{CDCVCT}
\end{eqnarray}
More details on how to implement new observables in SPheno can be found in Ref.\cite{Flavor2}.

\begin{table}[h]
\caption{Values used in the calculation.}
\begin{tabular}{@{}ccccccc@{}} \colrule
Vector&$\phi$&$J/\psi$&$\psi(2S)$&$\Upsilon(1S)$&$\Upsilon(2S)$&$\Upsilon(3S)$\\
\colrule
$m_V$&1.0194&3.0969&3.686&9.4603&10.023&10.355\\
$f_V$&0.241&0.418&0.294&0.649&0.481&0.539\\
BR($V\rightarrow ee$)&$2.973\times 10^{-4}$&$5.971\times 10^{-2}$&$7.93\times 10^{-3}$&$2.38\times 10^{-2}$&$1.91\times 10^{-2}$&$2.18\times 10^{-2}$\\
\colrule
\end{tabular}
\label{Tab2}
\end{table}
We present the numerical values of the vector masses, the decay constants and BR($V\rightarrow ee$) in Table \ref{Tab2}, which are taken from Ref.\cite{PDG2020} and Ref.\cite{Hazard2016}, and all the mass parameters are in GeV. Following the suggestion in Ref.\cite{Khodjamirian}, the transverse decay constants are set $f^T_V$=$f_V$ except for $J/\psi$, which has $f^T_{J/\psi}$ = 0.410GeV. Note that the branching ratio in Eq.(\ref{BR}) is largely independent of the values of the decay constants \cite{Hazard2016}.

The most accurate prediction for the mass of the W boson as well as the SM-like Higgs boson in the MRSSM is studied in Ref.\cite{Diessner2019}. A set of the benchmark point is given by \cite{Diessner2019}
\begin{equation}
\begin{array}{l}
\tan\beta=3,B_\mu=500^2,\lambda_d=1.0,\lambda_u=-0.8,\Lambda_d=-1.2,\Lambda_u=-1.1,\\
M_D^B=550,M_D^W=600,\mu_d=\mu_u=500,v_S=5.9,v_T=-0.38,\\
(m^2_l)_{11}=(m^2_l)_{22}=(m^2_l)_{33}=(m^2_r)_{11}
=(m^2_r)_{22}=(m^2_r)_{33}=1000^2,\\
(m^2_{\tilde{q}})_{11}=(m^2_{\tilde{u}})_{11}=(m^2_{\tilde{d}})_{11}=(m^2_{\tilde{q}})_{22}
=(m^2_{\tilde{u}})_{22}=(m^2_{\tilde{d}})_{22}=2500^2,\\
(m^2_{\tilde{q}})_{33}=(m^2_{\tilde{u}})_{33}=(m^2_{\tilde{d}})_{33}=1000^2,m_T=3000,m_S=2000.
\end{array}\label{N1}
\end{equation}
where all the mass parameters are in GeV or GeV$^2$. The predicted W boson mass in the MRSSM is comparable with the result from the combination of Large Electron-Positron collider and Fermilab Tevatron collider measurements \cite{CDF2013} and the result from the ATLAS collaboration \cite{ATLAS2018}. By changing the values of some parameters, e.g. $m_{SUSY}$, $v_T$, $\Lambda_u$ and $\Lambda_d$, the recent result on W boson mass from CDF collaboration \cite{CDF2022} can also be accommodated in the MRSSM. It is noted that these parameters have very small effect on the prediction of BR($V\rightarrow l_1 \bar{l}_2$) which take values along a narrow band. In the numerical analysis, the default values of the input parameters are set same with those in Eq.(\ref{N1}). The off-diagonal entries of the slepton mass matrices $m^2_l$, $m^2_r$ and the squark mass matrices $m^2_{\tilde{q}}$, $m^2_{\tilde{u}}$, $m^2_{\tilde{d}}$ in Eq.(\ref{N1}) are zero.

The LFV decays mainly originate from the off-diagonal entries of the soft breaking terms $m_{l}^{2}$ and $m_{r}^{2}$. These off-diagonal entries of $3\times3$ matrices $m_{l}^{2}$ and $m_{r}^{2}$ are parameterized by the mass insertions
\begin{eqnarray}
(m^{2}_{l})^{IJ}&=&\delta ^{IJ}_{l}\sqrt{(m^{2}_{l})^{II}(m^{2}_{l})^{JJ}},\nonumber\\
(m^{2}_{r})^{IJ}&=&\delta ^{IJ}_{r}\sqrt{(m^{2}_{r})^{II}(m^{2}_{r})^{JJ}},\nonumber
\end{eqnarray}
where $I,J=1,2,3$. To decrease the number of free parameters involved in our calculation, we assume $\delta ^{IJ}_{l}$ = $\delta ^{IJ}_{r}$ = $\delta ^{IJ}$ and then $\delta ^{IJ}$= $\delta ^{12}$, $\delta ^{13}$ or $\delta ^{23}$. The parameters $\delta^{IJ}$ are constrained by the experimental bounds on the LFV decays, such as the radiative two body decays $l_2\rightarrow l_1\gamma$, the leptonic three body decays $l_2\rightarrow 3l_1$ and $\mu-e$ conversion in nuclei. Current bounds of these LFV decays are listed in Table.\ref{Tab3} \cite{PDG2020}. 
\begin{table}[h]
\caption{Current bounds of $l_1\rightarrow l_2\gamma$, $l_1\rightarrow 3 l_2$ and $\mu-e$ conversion for a Ti target. }
\begin{tabular}{@{}cccccc@{}} \colrule
Decay&Bound&Decay&Bound&Decay&Bound\\
\colrule
$\mu\rightarrow e\gamma$&$4.2\times 10^{-13}$&$\tau\rightarrow e\gamma$&$3.3\times 10^{-8}$&$\tau\rightarrow \mu\gamma$&$4.4\times 10^{-8}$\\
$\mu\rightarrow 3e$&$1.0\times 10^{-12}$&$\tau\rightarrow 3e$&$2.7\times 10^{-8}$&$\tau\rightarrow 3\mu$&$2.1\times 10^{-8}$\\
$\mu-e$,Ti&$4.3\times 10^{-12}$&&&&\\\colrule
\end{tabular}
\label{Tab3}
\end{table}
In the following we will use these bounds to constrain the parameters $\delta^{IJ}$.

%%%%%%%%%%%%%%%%%%%%%%%%%%%%%%%%%%%%%%%%%%%%%%%%%%%%%%%%%%%%%%%%%%%
\begin{figure}[htbp]
\setlength{\unitlength}{1mm}
\centering
\begin{minipage}[c]{1\columnwidth}
\includegraphics[width=0.40\columnwidth]{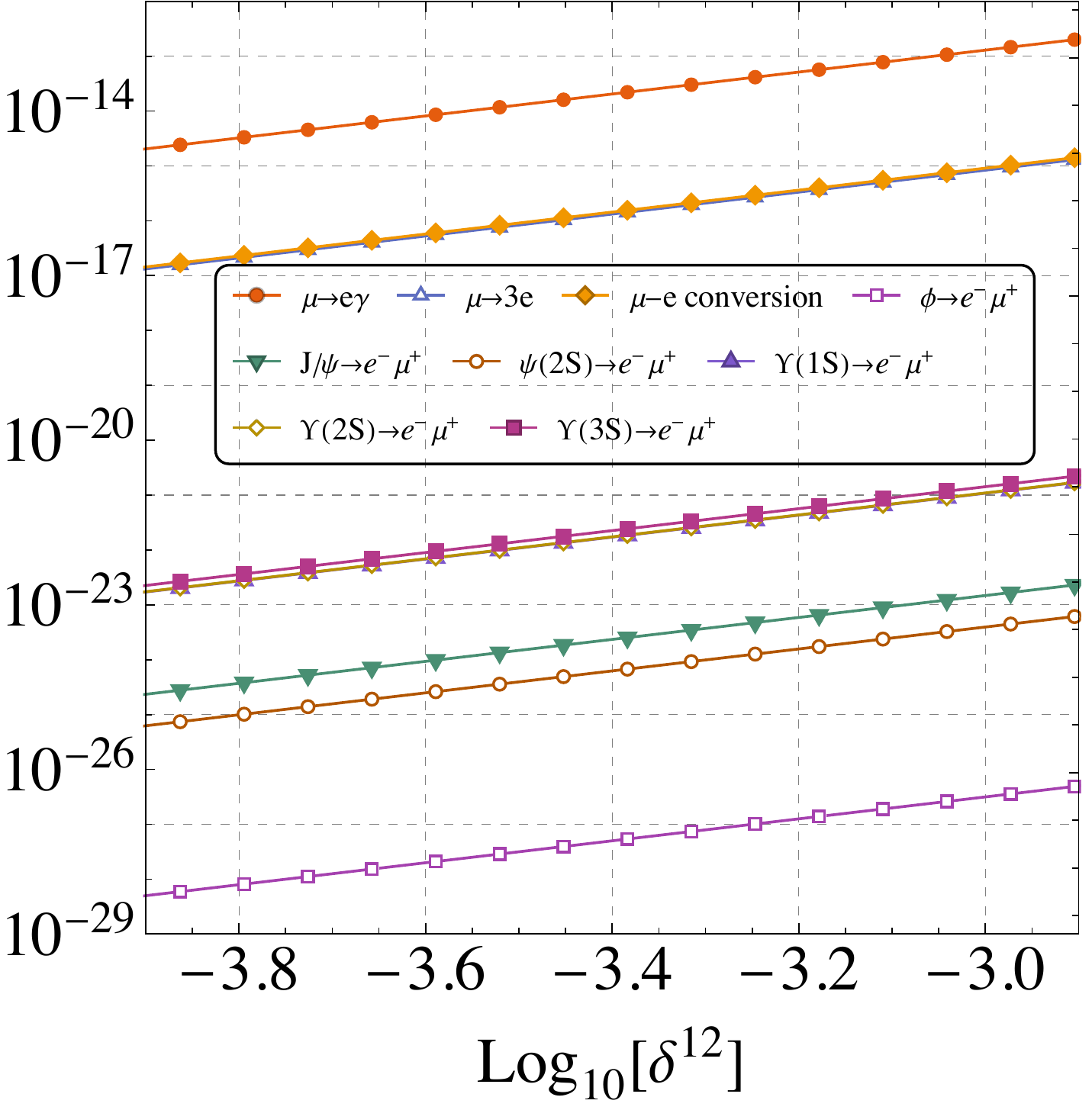}%
\includegraphics[width=0.40\columnwidth]{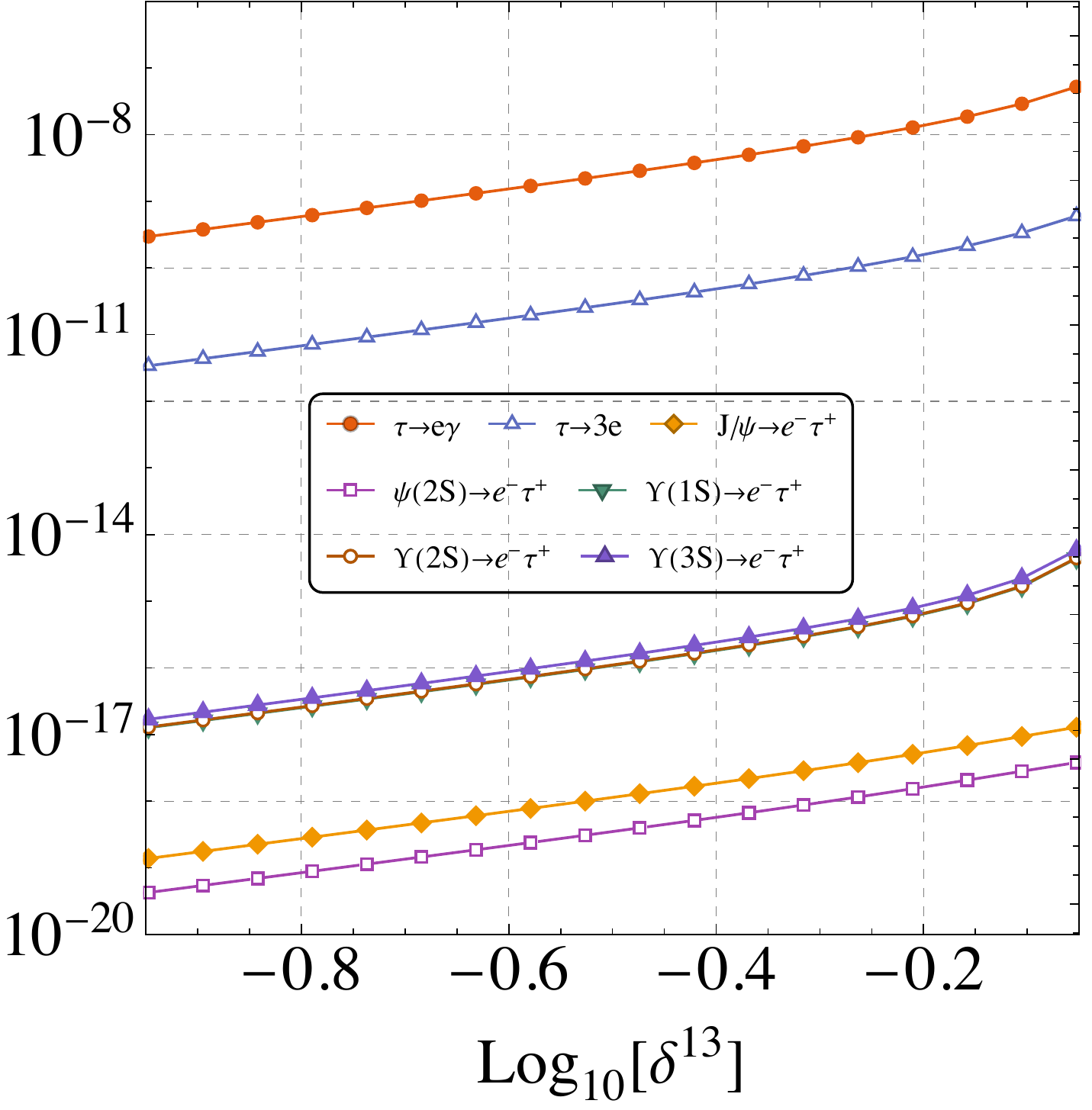}
\includegraphics[width=0.40\columnwidth]{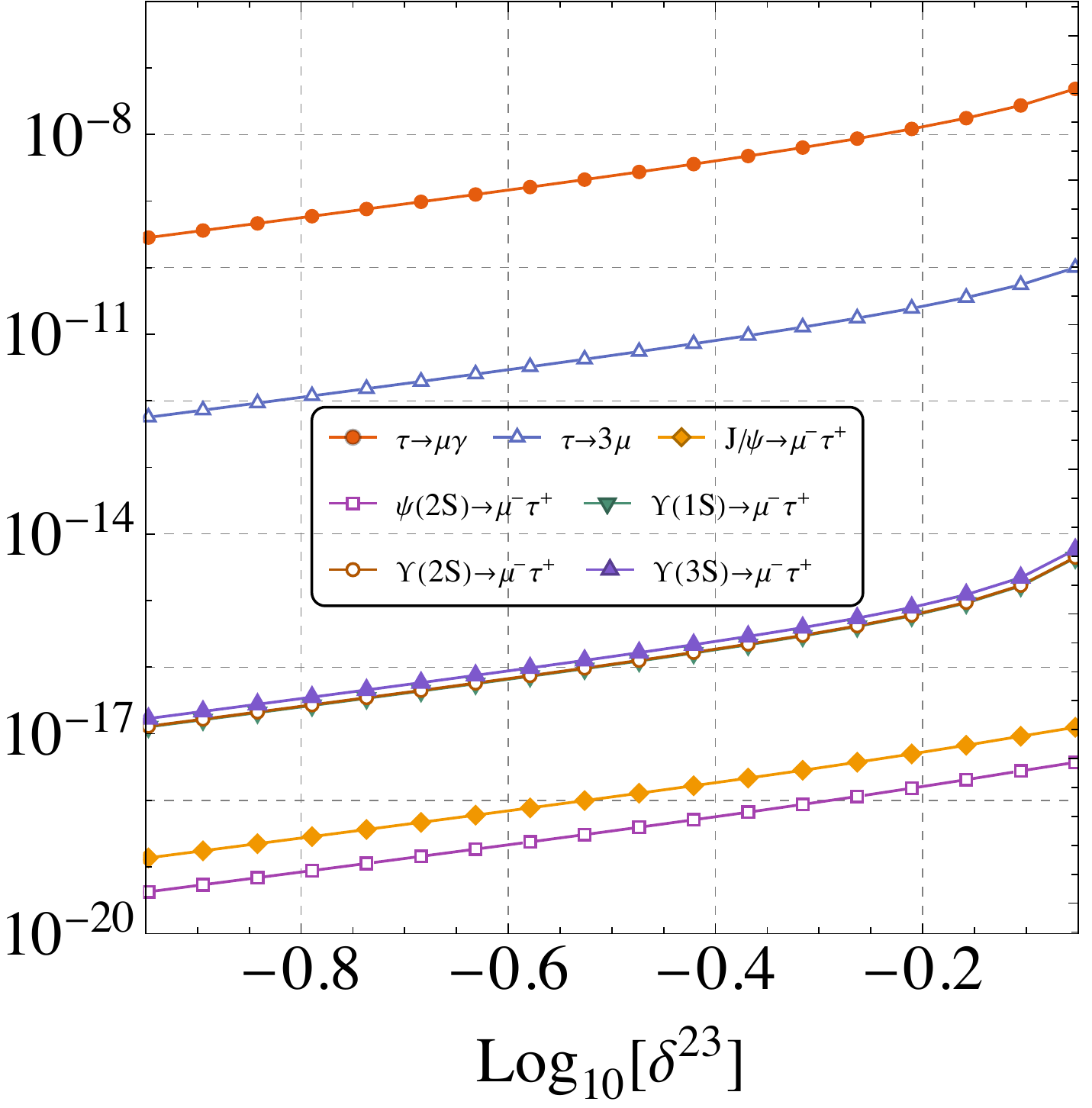}
\end{minipage}
\caption[]{Predictions for BR($V\rightarrow l_1 \bar{l}_2$) in the MRSSM. Note that the predictions for BR($\Upsilon(nS)\rightarrow l_1 \bar{l}_2$) in each plot are very close to each other.}
\label{figdata}
\end{figure}
%%%%%%%%%%%%%%%%%%%%%%%%%%%%%%%%%%%%%%%%%%%%%%%%%%%%%%%%%%%%%%%%%%%

We present the corresponding predictions for BR($V\rightarrow l_1 \bar{l}_2$) in Fig.\ref{figdata} along with the other SUSY parameters in Eq.(\ref{N1}) where the predictions for BR($l_1\rightarrow l_2\gamma$), BR($l_1\rightarrow 3 l_2$) and CR($\mu-e$, Ti) are also shown. Results are shown as a function of one of the parameters $\delta^{IJ}$. In all plots only the $\delta^{IJ}$ indicated is varied with all other mass insertions set to zero. The predictions for BR($\Upsilon(nS)\rightarrow l_1 \bar{l}_2$) in each plot are very close to each other. The prediction on BR($V\rightarrow e^-\mu^+$), BR($V\rightarrow e^-\tau^+$) and BR($V\rightarrow \mu^- \tau^+$) is affected by the mass insertions $\delta^{12}$, $\delta^{13}$ and $\delta^{23}$, respectively. The predictions on BR($V\rightarrow l_1 \bar{l}_2$) in the MRSSM are found to be below $10^{-14}$, which are at least seven orders of magnitude below the present experimental upper limits. A linear relationship in logarithmic scale is displayed between the branching ratios and the flavor violating parameters $\delta^{IJ}$. The actual dependence on $\delta^{IJ}$ is quadratic. The mentioned linear dependence is due to the fact that both x axis and y axis in Fig.\ref{figdata} are logarithmically scaled. In Fig.\ref{figdata} the following hierarchy is shown, BR($\Upsilon(3S)\rightarrow l_1 \bar{l}_2$)$>$BR($\Upsilon(2S)\rightarrow l_1 \bar{l}_2$)$\sim$BR($\Upsilon(1S)\rightarrow l_1 \bar{l}_2$)$>$BR($J/\psi\rightarrow l_1 \bar{l}_2$)$>$BR($\psi(2S)\rightarrow l_1 \bar{l}_2$)$>$BR($\phi\rightarrow l_1 \bar{l}_2$). The same hierarchy appears in several new physics \cite{Abada113013, Yue2016}. The most challenging experimental prospects for $\delta^{12}$ arise for $\mu\rightarrow e\gamma$. Considering the new sensitivity for BR($\mu\rightarrow e\gamma$) in the future projects will be about $6\times 10^{-14}$ from MEG II \cite{MEG2018}, $\delta^{12}$ is constrained to around $10^{-3}$. The constraints on $\delta^{13}$ from $\tau\rightarrow e\gamma$ and $\tau\rightarrow 3e$ are comparable and $\delta^{13}$ is constrained to around $10^{-0.2}$. The case for $\delta^{23}$ is same with $\delta^{13}$.

%%%%%%%%%%%%%%%%%%%%%%%%%%%%%%%%%%%%%%%%%%%%%%%%%%%%%%%%%%%%%%%%%%%
\begin{figure}[htbp]
\setlength{\unitlength}{1mm}
\centering
\begin{minipage}[c]{1\columnwidth}
\includegraphics[width=0.40\columnwidth]{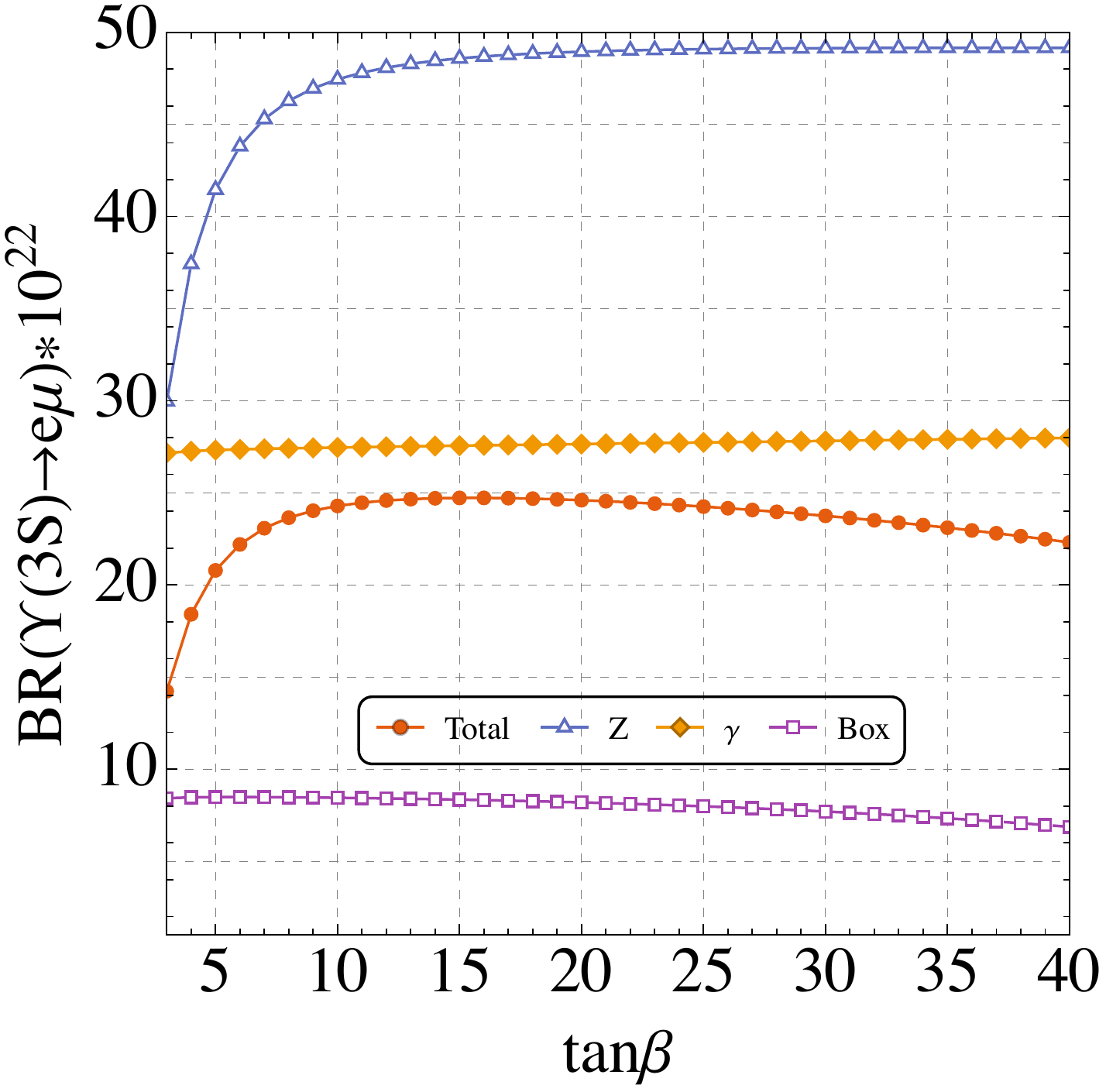}%
\includegraphics[width=0.40\columnwidth]{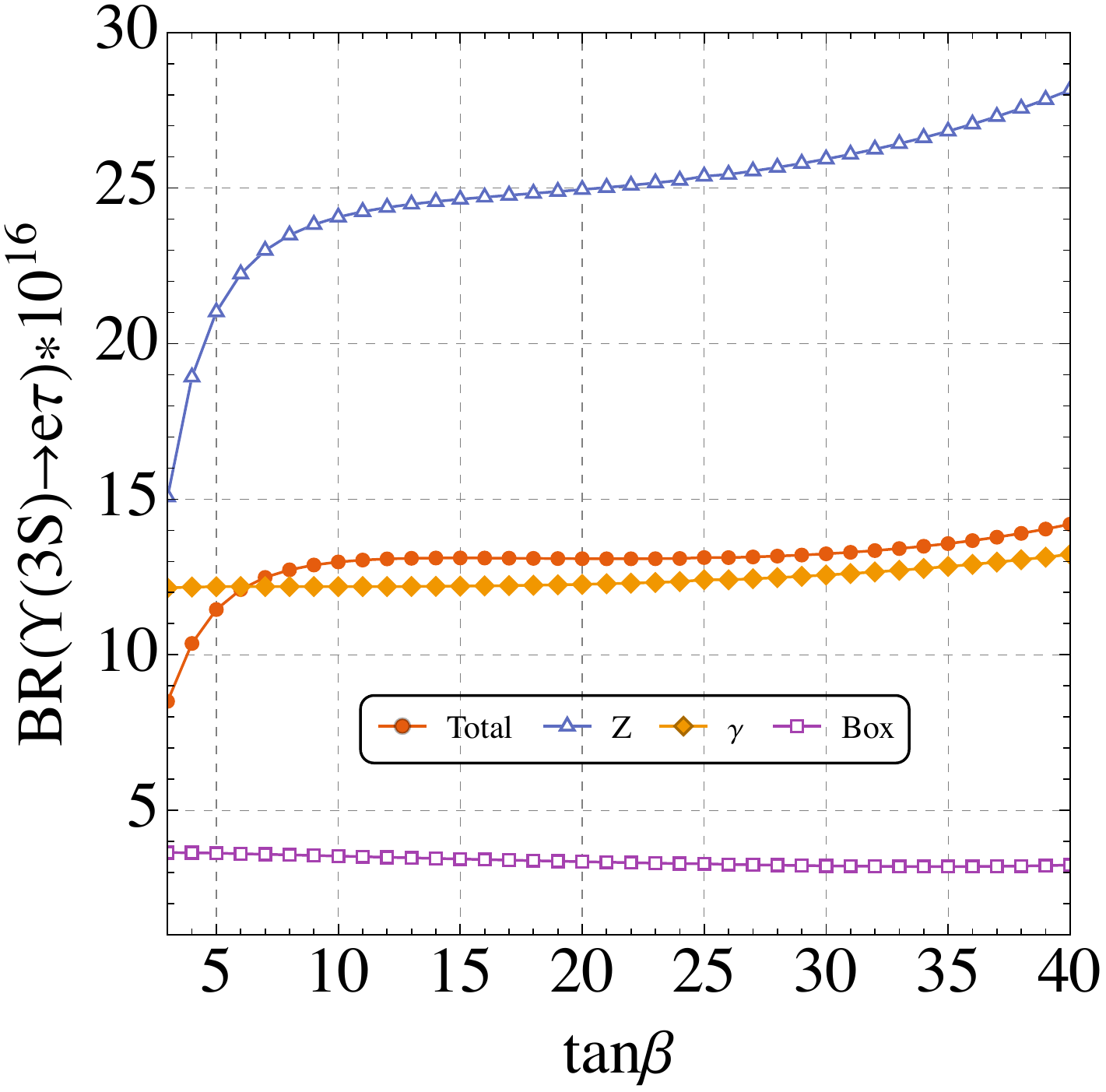}
\includegraphics[width=0.40\columnwidth]{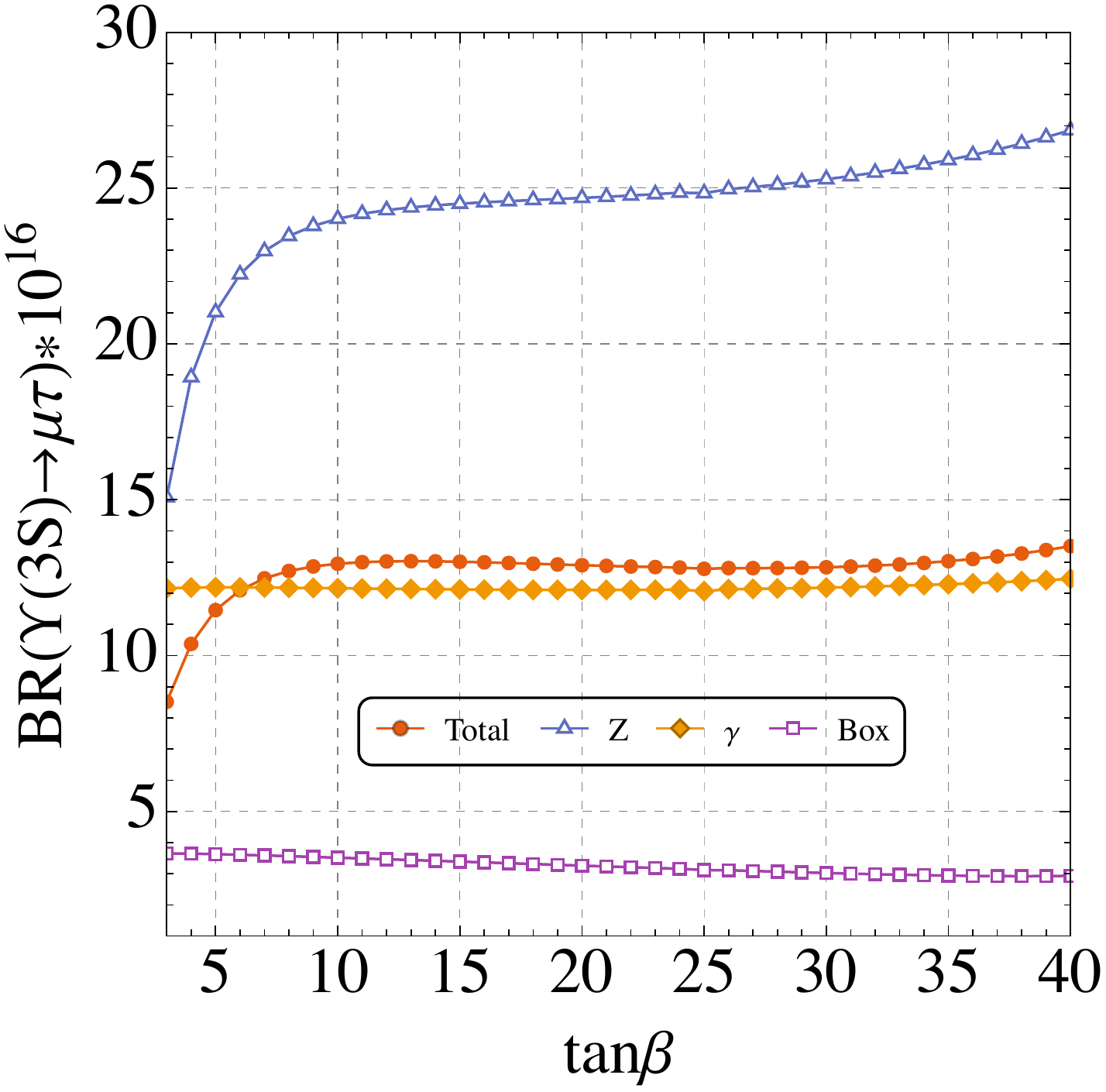}
\end{minipage}
\caption[]{Contributions to BR($\Upsilon(3S)\rightarrow l_1 \bar{l}_2$) from various parts in the MRSSM.}
\label{figY3}
\end{figure}
%%%%%%%%%%%%%%%%%%%%%%%%%%%%%%%%%%%%%%%%%%%%%%%%%%%%%%%%%%%%%%%%%%%

In the following, we will consider the process $\Upsilon(3S)\rightarrow l_1 \bar{l}_2$ as an example since the behavior for $\Upsilon(3S)$ also exists for $\Upsilon(1S)$, $\Upsilon(2S)$, $\phi$, $J/\psi$ and $\psi(2S)$. We present the corresponding predictions for BR($\Upsilon(3S)\rightarrow l_1 \bar{l}_2$) from various parts as a function of tan$\beta$ in Fig.\ref{figY3}, where $\delta^{12}$=$10^{-3}$($e\mu$), $\delta^{13}$ = $10^{-0.2}$ ($e\tau$) and $\delta^{23}$=$10^{-0.2}$ ($\mu\tau$) are used. The lines corresponding to `Z', `$\gamma$' and `Box' indicate the values of BR($\Upsilon(3S)\rightarrow l_1 \bar{l}_2$) are given by only the listed contribution with all others set to zero. The total prediction for BR($\Upsilon(3S)\rightarrow l_1 \bar{l}_2$) is also indicated. We observe that the contributions from different parts are comparable and the following hierarchy BR($\Upsilon(3S)\rightarrow l_1 \bar{l}_2$,Z)$>$BR($\Upsilon(3S)\rightarrow l_1 \bar{l}_2$,Total)$>$BR($\Upsilon(3S)\rightarrow l_1 \bar{l}_2$,$\gamma$)$>$BR($J/\psi\rightarrow l_1 \bar{l}_2$,Box) (for $J/\psi$ and $\psi(2S)$, BR($\Upsilon(3S)\rightarrow l_1 \bar{l}_2$,$\gamma$)$>$BR($\Upsilon(3S)\rightarrow l_1 \bar{l}_2$,Total)). It shows that varying tan$\beta$ has very small effect on the prediction of BR($\Upsilon(3S)\rightarrow l_1 \bar{l}_2$), which takes values along a narrow band, and indicates the tan$\beta$-enhancement for BR($\Upsilon(3S)\rightarrow l_1 \bar{l}_2$) does not exist in the MRSSM.

%%%%%%%%%%%%%%%%%%%%%%%%%%%%%%%%%%%%%%%%%%%%%%%%%%%%%%%%%%%%%%%%%%%
\begin{figure}[htbp]
\setlength{\unitlength}{1mm}
\centering
\begin{minipage}[c]{1\columnwidth}
\includegraphics[width=0.40\columnwidth]{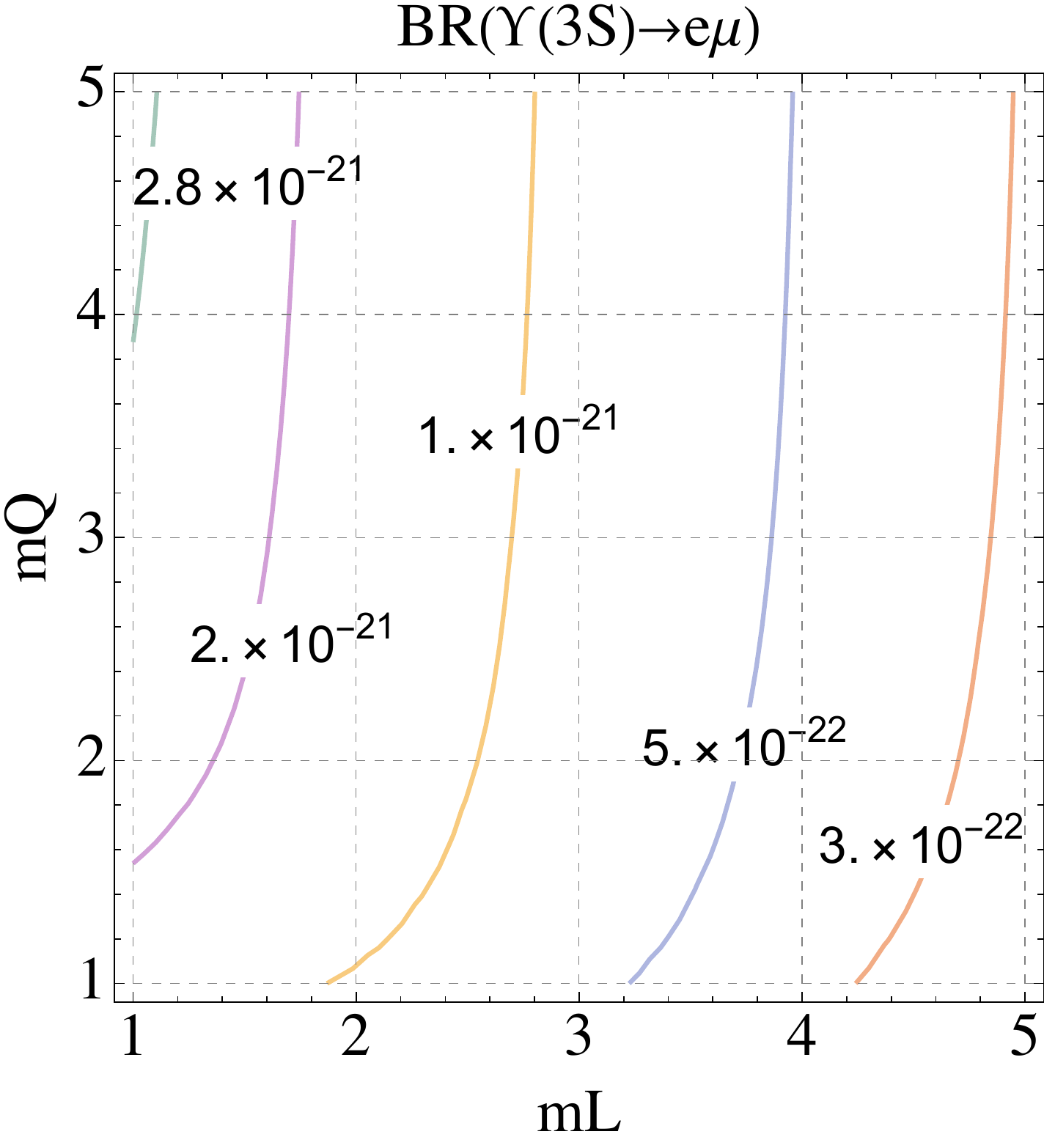}%
\includegraphics[width=0.40\columnwidth]{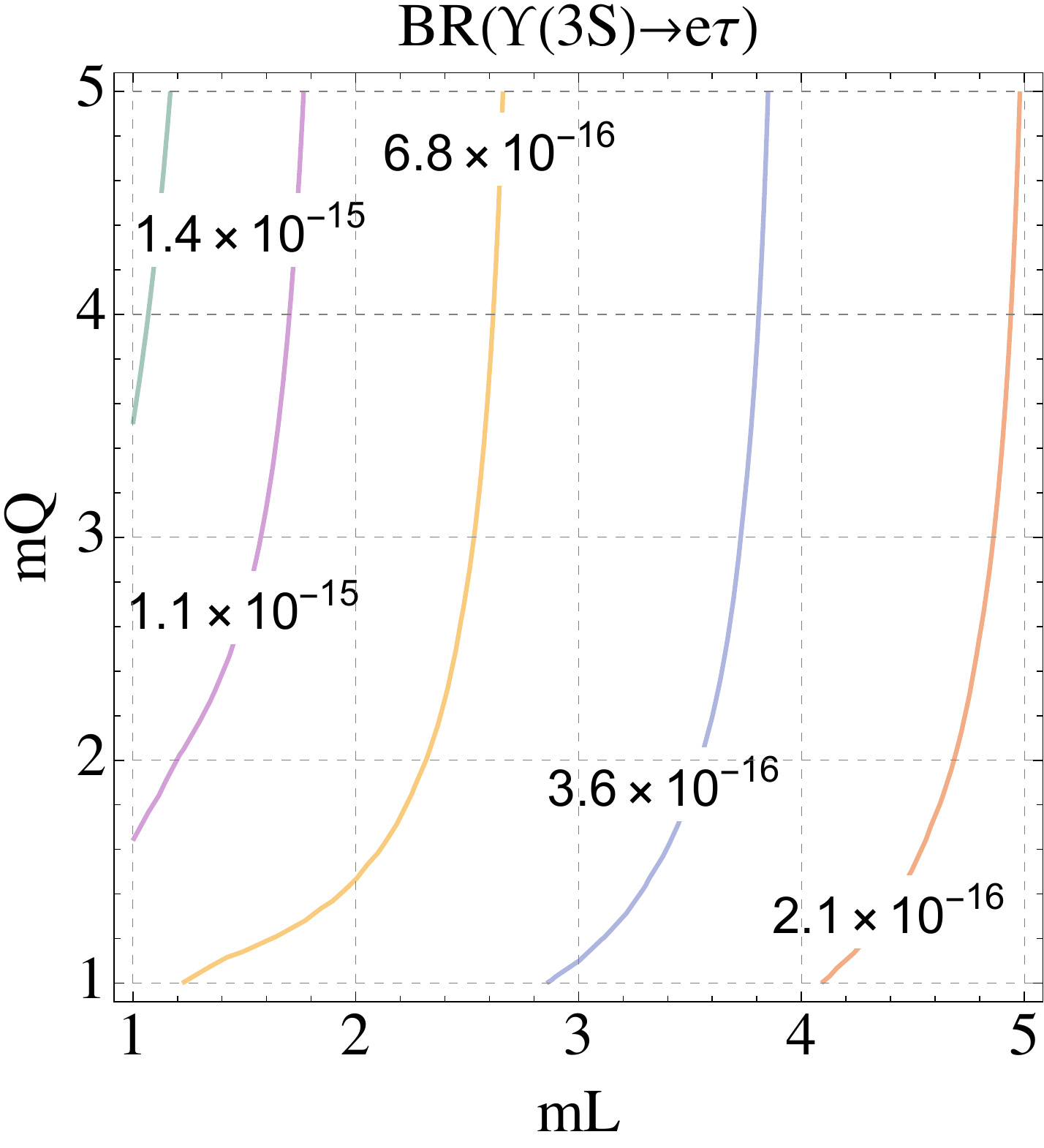}
\includegraphics[width=0.40\columnwidth]{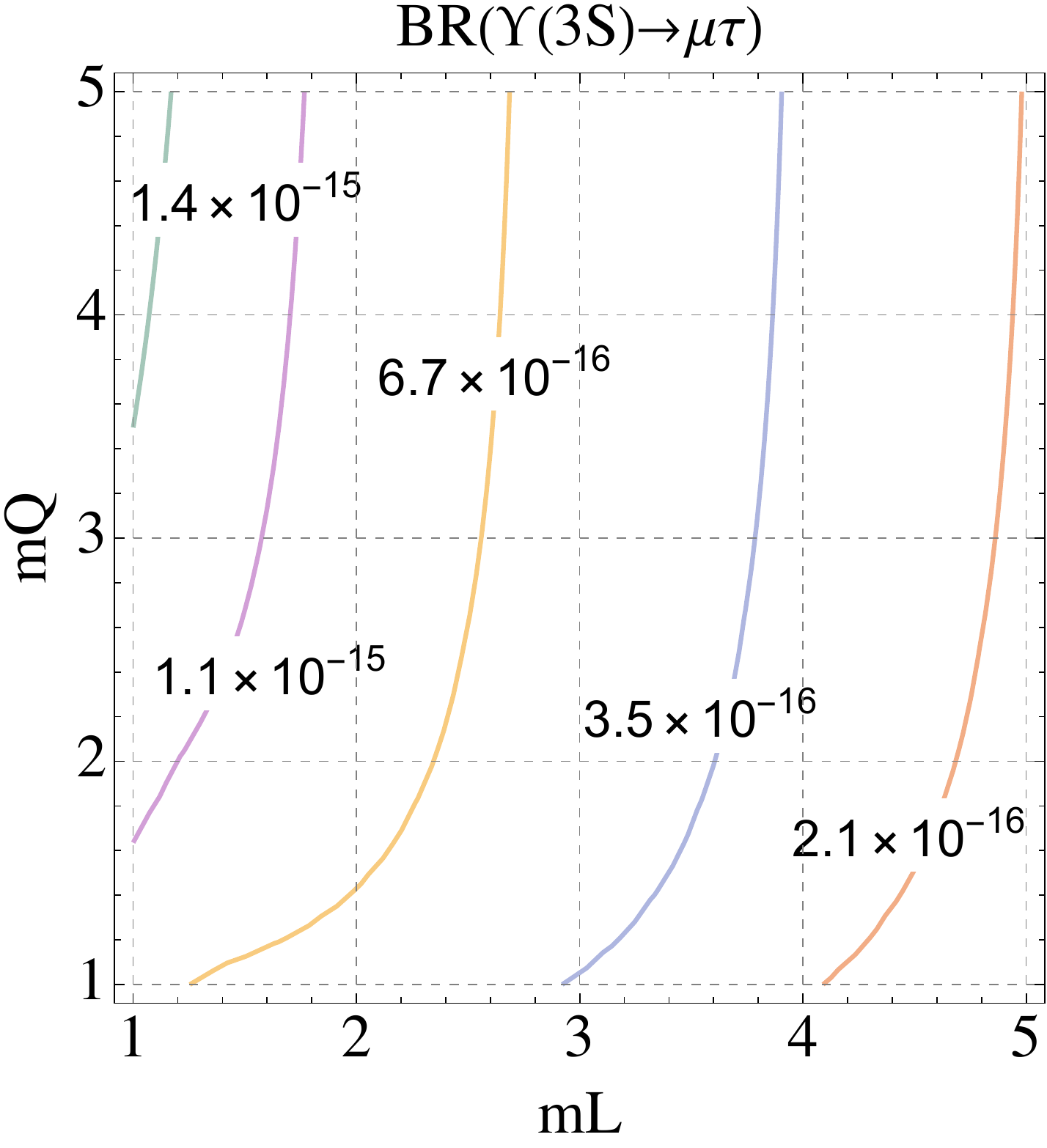}
\end{minipage}
\caption[]{Contour plots showing the behavior of BR($\Upsilon(3S)\rightarrow l_1 \bar{l}_2$) as functions of mQ and mL.}
\label{figY3lq}
\end{figure}
%%%%%%%%%%%%%%%%%%%%%%%%%%%%%%%%%%%%%%%%%%%%%%%%%%%%%%%%%%%%%%%%%%%

We present the contour plots of BR($\Upsilon(3S)\rightarrow l_1 \bar{l}_2$) in the mQ$\sim$mL plane in Fig.\ref{figY3lq}, where mQ=$\sqrt{(m^2_{\tilde{q}})_{ii}}=\sqrt{(m^2_{\tilde{u}})_{ii}}=\sqrt{(m^2_{\tilde{d}})_{ii}}$ ($i=1, 2, 3$) and mL=$\sqrt{(m^2_l)_{ii}}=\sqrt{(m^2_r)_{ii}}$ ($i=1, 2, 3$) are assumed. The predictions for BR($\Upsilon(3S)\rightarrow l_1 \bar{l}_2$) are sensitive to mQ and mL. The predictions on BR($\Upsilon(3S)\rightarrow l_1 \bar{l}_2$) increase slowly as mQ varies from 1 to 5 TeV and decrease slowly as mL varies from 1 to 5 TeV. In a wide region of mQ and mL, the predicted BR($\Upsilon(3S)\rightarrow l_1 \bar{l}_2$) can change about one order of magnitude. The off-diagonal entries $\delta^{IJ}_{\tilde{q},\tilde{u},\tilde{d}}$ of the squark mass matrices $m^2_{\tilde{q}}$, $m^2_{\tilde{u}}$ and $m^2_{\tilde{d}}$ have very small effect on the prediction of BR($\Upsilon(3S)\rightarrow l_1 \bar{l}_2$) which take values along a narrow band. The effect from the other parameters in Eq.(\ref{N1}) is same with that in Ref.\cite{kss2020,kss20202}.

The final results on the upper bounds of BR($V\rightarrow l_1 \bar{l}_2$) in the MRSSM are given in Table \ref{result}, which are obtained by assuming $\delta^{12}$=$10^{-3}$, $\delta^{13}$=$10^{-0.2}$($\approx 0.63$) and $\delta^{23}$=$10^{-0.2}$, respectively, where the results in the literature are also included for comparison. By means of an effective field theory, the data in Ref.\cite{Angelescu} are obtained from the recast of high-$p_T$ dilepton tails at the LHC for the left-handed scenario, where the dipole operators are not considered. The expressions for BR($V\rightarrow l_1 \bar{l}_2$) in Ref.\cite{Angelescu} are same except a few adjustments (e.g., mass, decay constant and full width). Since the mass, decay constant and full width for $\psi(2S)$ are very close to $J/\psi$, the bounds for $\psi(2S)\rightarrow l_1 \bar{l}_2$ are at the same level with $J/\psi\rightarrow l_1 \bar{l}_2$. For the same reason, the bounds for $\Upsilon(1S)\rightarrow l_1 \bar{l}_2$ and $\Upsilon(2S)\rightarrow l_1 \bar{l}_2$ are at the same level with $\Upsilon(3S)\rightarrow l_1 \bar{l}_2$. The predicted values for BR($V\rightarrow l \tau$) in the MRSSM lie a range between Ref.\cite{Abada113013} and Ref.\cite{Yue2016}, and the predicted values for BR($V\rightarrow e \mu$) are below Ref.\cite{Abada113013} and Ref.\cite{Yue2016}. All the direct bounds in new physics are smaller than the indirect bounds in Ref.\cite{Angelescu}.

\begin{table}[h]
\caption{ The upper bounds on BR($V\rightarrow l_1 \bar{l}_2$) in the literature and in the MRSSM.}
\begin{tabular}{@{}ccccc@{}} \colrule
Decay&Ref.\cite{Angelescu}(3 ab$^{-1}$)&Ref.\cite{Abada113013}((2,3)-ISS)&Ref.\cite{Yue2016}(331)&this work(MRSSM)\\
\colrule
$\phi\rightarrow e\mu$&$1.2\times 10^{-18}$&$1\times 10^{-23}$&$8.1\times 10^{-23}$&$3.2\times 10^{-27}$\\
$J/\psi\rightarrow e\mu$&$1.6\times 10^{-12}$&$2\times 10^{-20}$&$7.7\times 10^{-20}$&$1.5\times 10^{-23}$\\
$J/\psi\rightarrow e\tau$&$4.8\times 10^{-12}$&$1\times 10^{-19}$&$6.3\times 10^{-15}$&$5.3\times 10^{-18}$\\
$J/\psi\rightarrow \mu\tau$&$6.4\times 10^{-12}$&$4\times 10^{-19}$&$5.2\times 10^{-15}$&$5.3\times 10^{-18}$\\
$\psi(2S)\rightarrow e\mu$&-&$4\times 10^{-21}$&$2.1\times 10^{-20}$&$3.9\times 10^{-24}$\\
$\psi(2S)\rightarrow e\tau$&-&$4\times 10^{-20}$&$2.1\times 10^{-15}$&$1.6\times 10^{-18}$\\
$\psi(2S)\rightarrow \mu\tau$&-&$1\times 10^{-19}$&$1.7\times 10^{-15}$&$1.6\times 10^{-18}$\\
$\Upsilon(1S)\rightarrow e\mu$&-&$2\times 10^{-19}$&$3.8\times 10^{-17}$&$1.0\times 10^{-21}$\\
$\Upsilon(1S)\rightarrow e\tau$&-&$6\times 10^{-18}$&$5.5\times 10^{-12}$&$6.4\times 10^{-16}$\\
$\Upsilon(1S)\rightarrow \mu\tau$&-&$1\times 10^{-17}$&$4.3\times 10^{-12}$&$6.4\times 10^{-16}$\\
$\Upsilon(2S)\rightarrow e\mu$&-&$2\times 10^{-19}$&$4.2\times 10^{-17}$&$1.1\times 10^{-21}$\\
$\Upsilon(2S)\rightarrow e\tau$&-&$8\times 10^{-18}$&$6.1\times 10^{-12}$&$6.5\times 10^{-16}$\\
$\Upsilon(2S)\rightarrow \mu\tau$&-&$2\times 10^{-17}$&$4.8\times 10^{-12}$&$6.5\times 10^{-16}$\\
$\Upsilon(3S)\rightarrow e\mu$&$1.3\times 10^{-9}$&$5\times 10^{-19}$&$9.1\times 10^{-17}$&$1.4\times 10^{-21}$\\
$\Upsilon(3S)\rightarrow e\tau$&$7.9\times 10^{-9}$&$2\times 10^{-17}$&$1.3\times 10^{-11}$&$8.5\times 10^{-16}$\\
$\Upsilon(3S)\rightarrow \mu\tau$&$1.2\times 10^{-8}$&$3\times 10^{-17}$&$1.0\times 10^{-11}$&$8.5\times 10^{-16}$\\
\colrule
\end{tabular}
\label{result}
\end{table}

It is shown in Table \ref{result} that the bound on BR($J/\psi\rightarrow e^- \tau^+$) in the MRSSM is about $5.3\times 10^{-18}$ and this is ten orders of magnitude below the recently reported bound from the BESIII collaboration \cite{BES2103}. The bound on BR($J/\psi\rightarrow \mu^- \tau^+$) in the MRSSM is about $5.3\times 10^{-18}$ and this is twelve orders of magnitude below the current bound and ten orders of magnitude below the future experimental sensitivity ($1.5\times 10^{-8}$) \cite{BESIII2020}. The bound on BR($J/\psi\rightarrow e^-\mu^+$) in the MRSSM is about $1.5\times 10^{-23}$ and this is far below the current bound and the future experimental sensitivities ($6.0\times 10^{-9}$) \cite{BESIII2020}. The bound on BR($\Upsilon(nS)\rightarrow l \tau^+$) in the MRSSM is about ${\cal O} (10^{-16})$ and this is ten orders of magnitude below the recently reported bound from the Belle Collaboration \cite{Belle2022}. The bound on BR($\Upsilon(nS)\rightarrow e^- \mu^+$) in the MRSSM is about ${\cal O} (10^{-21})$ and this is fourteen orders of magnitude below the recently reported bound from the Belle Collaboration \cite{Belle2022} and the BaBar collaboration \cite{BABAR2022}.

\section{Conclusions\label{sec4}}

Although the higher order LFV processes in the SM are permitted, these are extremely suppressed by powers of the small neutrino masses and are not observable in current or planned experiments. Therefore, observation of the LFV decays would be a clear signature of new physics. In this work, we analyze the LFV decays of vector mesons $V\rightarrow l_1 \bar{l}_2$ in the MRSSM, by taking account of the constraints on the parameter space from the radiative charged lepton decays $l_1\rightarrow l_2\gamma$, leptonic three body decays $l_1\rightarrow 3l_2$ and $\mu-e$ conversion in nuclei. The prediction on BR($V\rightarrow e^-\mu^+$), BR($V\rightarrow e^-\tau^+$) and BR($V\rightarrow \mu^- \tau^+$) is affected by the mass insertions $\delta^{12}$, $\delta^{13}$and $\delta^{23}$, respectively. The final results on the upper bounds of BR($V\rightarrow l_1 \bar{l}_2$) in the MRSSM are given in Table \ref{result}, which are obtained by assuming $\delta^{12}$=$10^{-3}$, $\delta^{13}$=$10^{-0.2}$ and $\delta^{23}$=$10^{-0.2}$, respectively, where the results in the literature are also included for comparison. The predictions on BR($V\rightarrow l_1 \bar{l}_2$) in the MRSSM are far below the current upper limits. Thus, the LFV decays $V\rightarrow l_1 \bar{l}_2$ may be out reach of the near future experiments.

The studies of the radiative lepton flavor violating (RLFV) decays of vector mesons $V\rightarrow \gamma l_1 \bar{l}_2$ could provide important complementary access to search of new physics \cite{Hazard2016}. Besides the dipole, vector and tensor operators, the RLFV decays $V\rightarrow \gamma l_1 \bar{l}_2$ could receive contributions from the axial, scalar and pseudoscalar operators which are not accessible in $V\rightarrow l_1 \bar{l}_2$, e.g., the Higgs mediated self-energies and penguin diagrams. It might be possible that the RLFV processes $V\rightarrow \gamma l_1 \bar{l}_2$ can be enhanced close to the current or future experimental sensitivities while the LFV processes $V\rightarrow l_1 \bar{l}_2$ are still out reach of the current experiments.

\begin{acknowledgments}
\indent\indent
The work has been supported by the National Natural Science Foundation of China (NNSFC) under Grants No.11905002 and No.11805140, the Shanxi Scholarship Council of China (2021-31), the youth top-notch talent support program of the Hebei Province, the Foundation of Baoding University under Grant No. 2018Z01.

\end{acknowledgments}
%\appendix
%\section{}
%\label{appa}

\end{document}